\documentclass[11pt]{article}

\usepackage{amssymb}
\usepackage{graphicx}
\usepackage{layout}

\begin{document}

\title{A direct kinematical derivation of the relativistic Sagnac effect for light
or matter beams}
\author{Guido Rizzi$^{\S,\P}$ and Matteo Luca Ruggiero$^{\S,\P}$ \\
{\small $^\S$ Dipartimento di Fisica, Politecnico di Torino,}\\
{\small $^\P$ INFN, Sezione di Torino}\\
{\small E-mail guido.rizzi@polito.it, matteo.ruggiero@polito.it}}
\maketitle

\begin{abstract}
The Sagnac time delay and the corresponding Sagnac phase shift,
for relativistic matter and electromagnetic beams
counter-propagating in a rotating interferometer, are deduced on
the ground of relativistic kinematics. This purely kinematical
approach allows to explain the ''universality'' of the effect,
namely the fact that the Sagnac time difference does not depend on
the physical nature of the interfering beams. The only prime
requirement is that the counter-propagating beams have the same
velocity with respect to any Einstein synchronized local co-moving
inertial frame.
\end{abstract}

\section{Introduction}\label{sec:intro}

The phase shift due to the interference of two coherent light
beams, propagating in the two opposite directions along the rim of
a rotating ring interferometer, was observed for the first time by
Sagnac\cite{sagnac13} in 1913. Indeed, some years
before\cite{sagnac05}, he had predicted the following fringe shift
(with respect to the interference pattern when the device is at
rest), for monochromatic light waves in vacuum:

\begin{equation}
\Delta z=\frac{4\mathbf{\Omega \cdot S}}{\lambda c}
\label{eq:sagnac1}
\end{equation}
where $\mathbf{\Omega }$ is the (constant) angular velocity vector
of the turntable, $\mathbf{S}$ is the vector associated to the
area enclosed by the light path, and $\lambda $ is the wavelength
of light, as seen by an observer at rest on the rotating platform.
The time difference associated to the fringe shift
(\ref{eq:sagnac1}) turns out to be
\begin{equation}
\Delta t=\frac{\lambda }{c}\Delta z=\frac{4\mathbf{\Omega \cdot
S}}{c^{2}} \label{eq:sagnac2}
\end{equation}

His interpretation of these results was entirely in the framework
of the classical (non Lorentz!) ether theory; however, Sagnac was
the first scientist who reported an experimental observation of
the effect of rotation on space-time, which, after him, was named
''Sagnac effect''. It is interesting to notice that the Sagnac
effect was interpreted as a disproval of the Special Theory of
Relativity (SRT) not only during the early years of relativity (in
particular by Sagnac himself), but, also, more recently, in the
1990's by Selleri\cite{selleri96},\cite{selleri97},
Croca-Selleri\cite {croca99}, Goy-Selleri\cite{goy97},
Vigier\cite{vigier97}, Anastasovski et al.\cite{anastasowski99},
Klauber\cite{klauber}. However, this claim is incorrect: as a
matter of fact, the Sagnac effect for counter-propagating light
beams (in vacuum) can be explained completely in the framework of
SRT, see for instance Weber\cite{weber97}, Dieks\cite{dieks91},
Anandan\cite {anandan81}, Rizzi-Tartaglia\cite{rizzi98},
Bergia-Guidone \cite{bergia98},
Rodrigues-Sharif\cite{rodrigues01}, Henrisken\cite{henrisken86},
Rizzi-Ruggiero\cite{rizzi03}. According to SRT, eq.
(\ref{eq:sagnac2}) turns out to be just a first order
approximation of the relativistic proper time difference between
counter-propagating light beams.\bigskip

The experimental data show that the Sagnac fringe shift
(\ref{eq:sagnac1}) does not depend either on the light wavelength
nor on the presence of a co-moving optical medium. This is a first
important clue for the so called "universality of the Sagnac
effect". However, the most compelling claim for
the universal character of the Sagnac effect comes from the validity of eq. (%
\ref{eq:sagnac1}) not only for light beams, but also for any kind
of ''entities'' (such as electromagnetic and acoustic waves,
classical particles and electron Cooper pairs, neutron beams and
De Broglie waves and so on...) travelling in opposite directions
along a closed path in a rotating interferometer, with the same
(in absolute value) velocity with respect to the turntable. Of
course the entities take different times for a complete
round-trip, depending on their velocity relative to the turntable;
\textit{but the difference between these times is always given by
eq. (\ref {eq:sagnac2}).} So, the amount of the time difference is
always the same, both for matter and light waves, independently of
the physical nature of the interfering beams.

There have been many tests of the effect that prove its
universality. For instance, the Sagnac effect with matter waves
has been verified
experimentally using Cooper pair\cite{zimmermann65} in 1965, using neutrons%
\cite{atwood84} in 1984, using $^{40}Ca$ atoms
beams\cite{riehle91} in 1991 and using electrons, by
Hasselbach-Nicklaus\cite{hasselbach93}, in 1993. The effect of the
terrestrial rotation on neutrons phase was demonstrated in 1979 by
Werner et al.\cite{werner79} in a series of famous
experiments.\bigskip

However, as far as we know, a clear - and universally shared -
derivation of the Sagnac effect for matter waves, in the full
framework of SRT, seems to be lacking\footnote{As we pointed out
before,  the Sagnac effect has been derived by many Authors, in
the full framework of SRT, only for electromagnetic waves in
vacuum.} - or it is at least hard to find it in the literature.

In this paper we are going to provide a direct and simple
derivation of the Sagnac effect, using the relativistic law of
addition of velocities. Our derivation applies to any kind of
light or matter beams, counter-propagating in a rotating
interferometer. More explicitly, we shall show that, if a very
simple and sound requirement is fulfilled, the Sagnac time delay
does not depend on the physical nature of the interfering beams.
The ''simple and sound requirement'' is the following: the
counter-propagating beams must have the same velocity with respect
to any local co-moving inertial frame (LCIF), provided that it is
Einstein synchronized. Of course an alternative synchronization is
allowed\footnote{ Let us remind that an alternative
synchronization is actually needed \textit{globally} on the
platform.}, but this statement explicitly requires \textit{local}
Einstein's synchronization on the platform.

\section{The Sagnac Effect for material and light beams}\label{sec:direct}

\begin{figure}[here]
\begin{center}
\includegraphics[width=8cm,height=8cm]{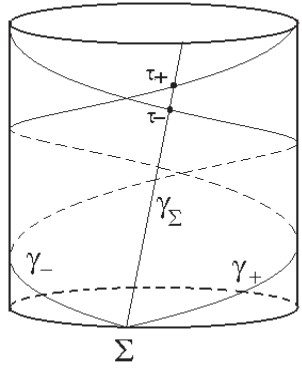}
\end{center}
\caption{{\protect\small The world-line of $\Sigma$, a point on
the disk
where a beam source and interferometric detector are lodged, is $%
\gamma_\Sigma$; $\gamma_+$ and $\gamma_-$ are the world-lines of
the co-propagating (+) and counter-propagating (-) beams. The
first intersection of $\gamma _{+}$ ($\gamma _{-}$) with $\gamma
_{\Sigma }$ takes place at the time $\tau_+$ ($\tau_-$), as
measured by a clock at rest in $\Sigma$.}} \label{fig:helix}
\end{figure}

Two light or matter beams are constrained to follow a circular
path along the rim of a rotating disk, with constant angular
velocity, in opposite directions. Let us suppose that a beam
source and an interferometric detector are lodged on a point
$\Sigma $ of the rim of the disk. Let $K$ be the central inertial
frame, parameterized by a set of cylindrical coordinates $\left\{
x^{\mu }\right\} ~=~\left( ct,r,\theta ,z\right) $, with line
element given by\footnote{ The signature is (-1,1,1,1), Greek
indices run from 0 to 3, while Latin indices run from 1 to 3.}

\begin{equation}
\mathrm{d}s^{2}=g_{\mu \nu }dx^{\mu }dx^{\nu }=-c^{2}\mathrm{d}t^{2}+\mathrm{%
d}r^{2}+r^{2}\mathrm{d}\theta ^{2}+\mathrm{d}z^{2}\mathrm{\ }
\label{eq:metricapiatta}
\end{equation}
In particular, if we confine ourselves to a disk ($z=const$), the
metric which we have to deal with is
\begin{equation}
ds^2=-c^2dt^2+dr^2+r^2d\theta^2  \label{eq:metrica}
\end{equation}

With respect to $K$, the disk (whose radius is $R$) rotates
with angular velocity $\Omega $, and the world line $\gamma _{\Sigma }$ of $%
\Sigma $ is
\begin{equation}
\gamma _{\Sigma }\equiv \left\{
\begin{array}{rcl}
x^{0} & = & ct \\
x^{1} & = & r=R \\
x^{2} & = & \theta =\Omega t
\end{array}
\right.  \label{eq:gammasigma0}
\end{equation}
or, eliminating $t$
\begin{equation}
\gamma _{\Sigma }\equiv \left\{
\begin{array}{rcl}
x^{0} & = & \frac{c}{\Omega }\theta \\
x^{1} & = & R \\
x^{2} & = & \theta
\end{array}
\right.  \label{eq:gammasigma}
\end{equation}

The world-lines of the co-propagating ($+$) and
counter-propagating ($-$) beams emitted by the source at time $%
t=0$ (when $\theta =0$) are, respectively:
\begin{equation}
\gamma _{+}\equiv \left\{
\begin{array}{rcl}
x^{0} & = & \frac{c}{\omega _{+}}\theta \\
x^{1} & = & R \\
x^{2} & = & \theta
\end{array}
\right.  \label{eq:gammapiu}
\end{equation}
\begin{equation}
\gamma _{-}\equiv \left\{
\begin{array}{rcl}
x^{0} & = & \frac{c}{\omega _{-}}\theta \\
x^{1} & = & R \\
x^{2} & = & \theta
\end{array}
\right.  \label{eq:gammameno}
\end{equation}
where $\omega _{+},\omega _{-}$ are their angular velocities, as
seen in the central inertial frame\footnote{Notice that $\omega
_{-}$ is positive if $|\omega _{-}^{\prime }|<\Omega $, null if
$|\omega _{-}^{\prime }|=\Omega $, and negative if $|\omega
_{-}^{\prime }|>\Omega $, see eq.(\ref{eq:betaunobeta}) below.}.
The first intersection of $\gamma _{+}$ ($\gamma _{-}$) with
$\gamma _{\Sigma }$ is the event ''absorption of the
co-propagating (counter-propagating) beam after a complete round
trip'' (see figure \ref{fig:helix}). This event takes place when
\begin{equation}
\frac{1}{\Omega }\theta _{\pm }=\frac{1}{\omega _{\pm }}(\theta
_{\pm }\pm 2\pi )  \label{eq:round}
\end{equation}
where the $+$ ($-$) sign holds for the co-propagating
(counter-propagating) beam. The solution of eq. (\ref{eq:round})
is:
\begin{equation}
\theta _{\pm }=\pm \frac{2\pi \Omega }{\omega _{\pm }-\Omega }
\label{eq:thetapiu}
\end{equation}
If we introduce the dimensionless velocities $\beta =\Omega R/c$,
$\beta _{\pm }=\omega _{\pm }R/c$, the $\theta $-coordinate of the
absorption event can be written as follows:
\begin{equation}
\theta _{\pm }=\pm \frac{2\pi \beta }{\beta _{\pm }-\beta }
\label{eq:thetapiumeno}
\end{equation}
The proper time read by a clock at rest in $\Sigma $ is given by
\begin{equation}
\tau =\frac{1}{c}\int_{\gamma _{\Sigma
}}ds=\frac{1}{c}\int_{\gamma _{\Sigma
}}\sqrt{c^{2}dt^{2}-R^{2}d\theta ^{2}}=\frac{1}{\Omega }\sqrt{1-\beta ^{2}}%
\int_{\gamma _{\Sigma }}d\theta  \label{eq:dtau}
\end{equation}
Taking into account eq. (\ref{eq:thetapiumeno}), the proper time
$\tau_+$ ($\tau_-$) elapsed between the emission and the
absorption of the co-propagating (counter-propagating) beam, read
by a clock at rest in $\Sigma $, is given by
\begin{equation}
\tau _{\pm }=\ \ \pm \frac{2\pi \beta }{\Omega }\frac{\sqrt{1-\beta ^{2}}}{%
\beta _{\pm }-\beta }  \label{eq:taupiumeno}
\end{equation}
and the proper time difference $\Delta \tau \equiv \tau _{+}-\tau
_{-}$ turns out to be
\begin{equation}
\Delta \tau =\frac{2\pi \beta }{\Omega }\sqrt{1-\beta
^{2}}\frac{\beta _{-}-2\beta +\beta _{+}}{(\beta _{+}-\beta
)(\beta _{-}-\beta )} \label{eq:deltatau11}
\end{equation}

Without specifying any further conditions, the proper time
difference (\ref{eq:deltatau11}) appears to depend upon $\beta
,\beta _{+},\beta _{-}$: this means that it does depend, in
general, both on the velocity of rotation of the disk and on the
velocities of the beams.

Let $\beta _{\pm }^{\prime }$ be the dimensionless velocities of
the beams as measured in any Minkowski inertial frame, locally
co-moving with the rim of the disk, or briefly speaking in any
locally co-moving inertial frame (LCIF). Provided that each LCIF
is Einstein synchronized (see Rizzi-Serafini
\cite{rizzi-serafini-libro}), the Lorentz law of velocity addition
gives the following relations between $\beta _{\pm }^{\prime }$
and $\beta _{\pm }$:
\begin{equation}
\beta _{\pm }=\frac{\beta _{\pm }^{\prime }+\beta }{1+\beta _{\pm
}^{\prime }\beta }  \label{eq:betaunobeta}
\end{equation}
By substituting (\ref{eq:betaunobeta}) in (\ref{eq:deltatau11}) we
easily obtain
\begin{equation}
\Delta \tau =\frac{4\pi \beta ^{2}}{\Omega }\frac{1}{\sqrt{1-\beta ^{2}}}+%
\frac{2\pi \beta }{\Omega }\frac{1}{\sqrt{1-\beta ^{2}}}\left( \frac{1}{%
\beta _{+}^{\prime }}+\frac{1}{\beta _{-}^{\prime }}\right)
\label{eq:deltatau12}
\end{equation}

Now, let us impose the condition ''equal relative velocity in
opposite directions'':
\begin{equation}
\beta _{+}^{\prime }=-\beta _{-}^{\prime }  \label{eq:lec}
\end{equation}

Such condition means that the beams are required to have the same
velocity (in absolute value) in every LCIF, provided that every
LCIF is Einstein synchronized. If condition (\ref{eq:lec}) is
imposed, the proper time difference (\ref{eq:deltatau12}) reduces
to
\begin{equation}
\Delta \tau =\frac{4\pi \beta ^{2}}{\Omega }\frac{1}{\sqrt{1-\beta ^{2}}}=%
\frac{4\pi R^{2}\Omega }{c^{2}}\left( 1-\frac{\Omega ^{2}R^{2}}{c^{2}}%
\right) ^{-1/2}  \label{eq:deltatausagnac}
\end{equation}
which is the relativistic Sagnac time difference.

A very relevant conclusion follows. According to eq. (\ref
{eq:taupiumeno}), the beams take different times - as measured by
the clock at rest on the starting-ending point $\Sigma $ on the
platform - for a complete round trip, depending on their
velocities $\beta _{\pm }^{\prime }$ relative to the turnable.
However, when condition (\ref{eq:lec}) is imposed, the difference
$\Delta \tau $ between these times does depend only on the angular
$\Omega $ of the disk, and it does not depend on the velocities of
propagation of the beams with respect the turnable.

This is a very general result, which has been obtained on the
ground of a purely kinematical approach. The Sagnac time
difference (\ref {eq:deltatausagnac}) applies to any couple of
(physical or even mathematical) entities, as long as a velocity,
with respect the turnable, can be consistently defined. In
particular, this result applies as well to photons (for which
$|\beta _{\pm }^{\prime }|=1$), and to any kind of classical or
quantum particle under the given conditions (or
electromagnetic/acoustic waves in presence of an homogeneous
co-moving medium)\footnote{ Provided that a group velocity can be
defined.}. This fact evidences, in a clear and straightforward
way, the universality of the Sagnac effect.

\section{A remark on the synchronization}\label{sec:remsync}

As it is well known (see for instance
Rizzi-Serafini\cite{rizzi-serafini-libro} or
Minguzzi\cite{minguzzi}), in a local or global inertial frame (IF)
the synchronization
can be arbitrarily chosen within the\textit{\ synchronization gauge }%
{\normalsize
\begin{equation}
\left\{
\begin{array}{l}
t^{\prime }=t^{\prime }\,(\,t,\,x^{1},x^{2},x^{3}) \\
x_{i}^{\prime }=x_{i}
\end{array}
\right.   \label{eq:synchr}
\end{equation}
}(with the additional condition $\partial t^{\prime }/\partial
t>0$, which ensures that the change of time parameterization does
not change the arrow of time)\footnote{Eq. (\ref{eq:synchr}) is a
subset of the set of all the possible parameterizations of the
given physical IF, see for instance Cattaneo\cite{cattaneo}, M\o ller\cite{moller}, Nikoli\'c\cite{nikolic}.}. In eq. (%
\ref{eq:synchr}) the coordinates ($t,x_{i}$) are Einstein coordinates, and ($%
t^{\prime },x_{i}^{\prime }$) are re-synchronized coordinates of
the IF under consideration. Of course, the IF turns out to be
optically isotropic
if and only if it is parameterized by Einstein coordinates ($t,x_{i}$).%
\footnote{We want to point out that    the local isotropy or
anisotropy of the velocity of light in an IF is not a fact, with a
well defined ontological meaning, but a convention which depends
on the synchronization chosen in the
IF\cite{rizzi-serafini-libro},\cite{minguzzi}. In particular, the
velocity of light has the invariant value $c$ in every LCIF, both
in co-rotating and counter-rotating direction, if and only if the
LCIF are Einstein-synchronized. }

According to the previous section, the central inertial frame $K$
is Einstein synchronized; let us call $F(K)$ the ''simultaneity
foliation'' of space-time with respect to $K$. However, on the
rotating platform many synchronization choices can be done,
depending on the aims and circumstances. In particular, exploiting
the gauge freedom, two different synchronization choices turn out
to be specially useful.

If we look for a global synchronization on the rotating platform,
any LCIF must share the synchronization of $K$, that is the
''simultaneity foliation'' $F(K)$ of space-time.

On the other hand, if  we look for a plain kinematical
relationship between local velocities, in order to explain the
universality of the Sagnac effect, any LCIF must be Einstein
synchronized. In fact, only Einstein's synchronization allows the
clear and meaningful requirement\footnote{Formally expressed by
condition (\ref{eq:lec}).}: ''equal relative velocity in opposite
directions''.

\section{A remark on the interferometric detectability of the Sagnac effect}\label{sec:direct_remark}

The Sagnac time difference (\ref{eq:deltatausagnac}) also applies
to the Fourier components of the wave packets associated to a
couple of matter beams counter-propagating, with the same relative
velocity, along the rim. Of course only matter beams are physical
entities, while Fourier components are just mathematical entities,
which no energy transport is associated to.

With regard to the interferometric detection of the Sagnac effect,
the crucial point is the following. Despite the lack of a direct
physical meaning and energy transfer, the phase velocity of these
Fourier components (which is the same for both the co-rotating and
counter-rotating ones) complies with the Lorentz law of velocity
composition (\ref{eq:betaunobeta}).

Moreover, the interferometric detection of the Sagnac effect
requires that the wave packet associated to the matter beam should
be sharp enough in the frequency space to allow the appearance, in
the interferometric region, of an observable fringe
shift\footnote{ That is, the Fourier components of the packet wave
should have slightly different wavelengths.}. It may be worth
recalling that:

 (i) the observable fringe shift $\Delta z$ depends on
to the phase velocity of the Fourier components of the packet
wave;

 (ii) with respect to an Einstein synchronized LCIF,
the velocity of every Fourier component of the wave packet
associated to the matter beam, moving with the velocity (in
absolute value) $v\equiv c|\beta
_{\pm }^{\prime }|$,\textbf{\ }is given by the De Broglie expression $%
v_{f}=c^{2}/v$.

 The consequent Sagnac phase shift, due to the
relativistic time difference (\ref{eq:deltatausagnac}), is
\begin{equation}
\Delta \Phi =2\pi \Delta z=2\pi \left( \frac{v_{f}}{\lambda
}\Delta \tau \right) =\frac{8\pi ^{2}R^{2}\Omega }{\lambda
v}\left( 1-\frac{\Omega ^{2}R^{2}}{c^{2}}\right) ^{-1/2}
\label{eq:phaseshift}
\end{equation}

\section{Conclusions}\label{sec:conclusions}

 We have given a direct derivation of the Sagnac
effect on the bases of the relativistic kinematics. In particular,
only the law of velocities addition, together with the condition
that the counter-propagating beams have the same velocity with
respect to any Einstein synchronized LCIF, have been used to
obtain the Sagnac time difference. In this way, we have shown,  in
a straightforward way, the independence of the Sagnac time
difference from the physical nature and the velocities (relative
to the turntable) of the interfering beams.

The simple derivation that we have outlined proves, in a clear and
understandable way, the universal features of the Sagnac effect,
which can be clearly understood as a purely geometrical effect in
the Minkowski space-time of SRT, while it would be hard to grasp
in the context of classical physics.

\end{document}